\documentclass[10pt, conference, compsocconf]{IEEEtran}
\usepackage[utf8]{inputenc}

\usepackage{graphicx}
\graphicspath{ {./fig/} }
\usepackage{float}
\usepackage{amsmath}
\usepackage{listings}
\usepackage{courier}

\usepackage{url}

\usepackage[table,x11names,dvipsnames,table]{xcolor}
\usepackage[colorinlistoftodos]{todonotes}

\usepackage[para]{threeparttable}

\usepackage[]{algorithm2e}
\usepackage{fancyvrb}

\begin{document}
\title{A low-overhead approach for\\ self-sovereign identity in IoT}

\author{
\IEEEauthorblockN{
Geovane Fedrecheski\IEEEauthorrefmark{1},
Laisa C. P. Costa\IEEEauthorrefmark{1},
Samira Afzal\IEEEauthorrefmark{1},
Jan M. Rabaey\IEEEauthorrefmark{4},
Roseli D. Lopes\IEEEauthorrefmark{1},
Marcelo K. Zuffo\IEEEauthorrefmark{1}}
\IEEEauthorblockA{
\IEEEauthorrefmark{1}Interdisciplinary Center on Interactive Technologies, Polytechnic School,\\
University of Sao Paulo, Brazil\\
\IEEEauthorrefmark{4}Berkeley Wireless Research Center, Electrical Engineering and Computer Science Department,\\
University California, Berkeley, US
}\\
\{geovane, laisa, mkzuffo\}@lsi.usp.br, jan\_rabaey@berkeley.edu
\thanks{This research was partially funded by CAPES and LSI-TEC.}
}

\IEEEoverridecommandlockouts


\maketitle

\begin{abstract}
We present a low-overhead mechanism for self-sovereign identification and communication of IoT agents in constrained networks.
Our main contribution is to enable native use of Decentralized Identifiers (DIDs) and DID-based secure communication on constrained networks, whereas previous works either did not consider the issue or relied on proxy-based architectures.
We propose a new extension to DIDs along with a more concise serialization method for DID metadata. Moreover, in order to reduce the security overhead over transmitted messages, we adopted a binary message envelope.
We implemented these proposals within the context of Swarm Computing, an approach for decentralized IoT. Results showed that our proposal reduces the size of identity metadata in almost four times and security overhead up to five times. We observed that both techniques are required to enable operation on constrained networks.
\end{abstract}
\IEEEpeerreviewmaketitle


\section{Introduction}
\label{sec:introduction}


\IEEEPARstart{S}{elf}-sovereign identity (SSI), also referred to as decentralized identity, is an emerging approach that enables subjects to be in full control of their own digital identities \cite{ferdous2019search}. When applied to IoT environments, SSI facilitates device ownership, enhances privacy, 
and reduces dependency on third parties \cite{fedrecheski2020self}.
IoT approaches that rely on decentralized architectures, such as the Swarm \cite{costa2015swarm_tce}, are expected to greatly benefit from these new capabilities enabled by SSI.

Once devices are put in charge of their own identity, new challenges arise, mainly due to the limitations of constrained devices and networks. 
In this paper, we focus on reducing the overhead of self-sovereign identity in IoT networks. 
We extend existing standards to reduce message footprint and propose a new serialization method that significantly reduces the transmitted bytes.

The current approach to implement self-sovereign identity relies on the use of Decentralized Identifiers (DIDs) \cite{W3C2019decentralized}. 
A DID is a new form of identifier that does not depend on trusted third parties and has an associated set of cryptographic metadata referred to as a DID Document (DDo). 
This way, beyond simple identification, a DDo enables the establishment of an end-to-end secure channel, which can be done using a transport-agnostic protocol called DIDComm \cite{DIF2020didcomm}.
The DID data model is extensible, and by March 2021 there are around 90 registered extensions (referred to as ``DID methods'') \cite{steele2019did}.

Most works on DIDs, however, overlook the overhead of transmitting DDos, a crucial aspect in bandwidth-constrained networks. For example, while the LoRa network only allows packets of up to 242 bytes, the most compact of the registered DID extensions requires DDos in the order of 500 bytes.
Even the works applying DIDs in the IoT context did not consider size limitations imposed by network bandwidth \cite{su2020secure} \cite{fotiou2019enabling}. In one approach that does consider resource constraints, DIDs transmission is avoided by using OAuth tokens in a centralized architecture \cite{lagutin2019enabling}.
Furthermore, the overhead on secure communications imposed by the DIDComm protocol also has not been addressed in the literature.

Considering the potential benefits of the self-sovereign approach for the IoT, and the drawbacks of existing solutions, this paper proposes a low-overhead method for DID-based identification and secure communication. 
The contributions of this paper are as follows:
\begin{itemize}
\item A new DID method suitable for IoT networks referred to as \textit{DID Swarm}, which has smaller DID and DDo sizes when compared to existing methods.
\item CBOR-based DID Documents for IoT (CBOR-DI), a novel serialization mechanism that can reduce DDo sizes by almost four times.
\item DIoTComm, a binary envelope to replace DIDComm in IoT networks that reduces overhead up to five times.
\item Integration with the Swarm framework, a decentralized IoT approach that enables spontaneous resource sharing.
\end{itemize}

In the remainder of the paper, we present the related work and some background, followed by the system design, the evaluation results, and our conclusions and next steps.



\section{Related Work}
\label{sec:related-work}

Several works have proposed identity solutions for the IoT, however, many of them require centralized management. The Open Connectivity Foundation \cite{open2020ocf} uses Fully Qualified Domain Names and relies on certificates for identity management, two centralized approaches. The Web of Things framework \cite{w3c2020web} relies on Uniform Resource Identifiers (URIs), which can be decentralized, however, the identity management is still done via certificates.

Our previous work shows the potential of DIDs as an owner-centric, privacy-aware and decentralized identification mechanism for IoT applications \cite{fedrecheski2020self}.
One of the challenges for DID adoption in the IoT, however, is communication overhead, since none of the currently registered DID methods \cite{steele2019did}  has been designed to work in constrained networks.
For example, the Sovrin DID method \cite{sovrin2020did}
uses approximately 500 bytes to encode a DID Document.

Existing works have applied self-sovereign identity to IoT. 
In one case \cite{su2020secure}, authors propose an architecture for machine identifiers based on DIDs, along with a storage layer based on Blockchain and IPFS. Another approach \cite{fotiou2019enabling} combines DIDs with Verifiable Credentials, a data model for signed attributes in SSI, to manage identification in the IoT. 
Others \cite{lagutin2019enabling} have used newly generated DIDs to populate access control lists and enable guests to access smart home devices.
None of these works, however, considered the size of the documents associated with implementation of SSI in low-power IoT networks. Moreover, it is not clear how these works protect the communication between agents.


To enable secure communications based on DIDs, the DIDComm protocol has been proposed \cite{DIF2020didcomm}. 
DIDComm supports authenticated message exchanges and message routing over loosely trusted routers, and is independent of transport protocol.
Nevertheless, DIDComm uses JSON for serialization, which implies an overhead prevents its use in low-power IoT networks. Currently there is no known low-overhead alternative to DIDComm.

\section{Background}
\label{sec:background}

\subsection{Self-Sovereign Identity}
In the SSI approach each entity has full control of its own identity. 
Formally, the complete self-sovereign identity of an agent is the union of all of its identifiers and attributes across different domains \cite{ferdous2019search}. 
The Decentralized Identifiers (DID) specification \cite{W3C2019decentralized} defines a new format for self-sovereign identifiers and related metadata.
A DID is composed of a DID method\footnote{A ``DID method'' is an extension of the DID specification;} prefix and a namespace-specific identifier (NSI) \cite{W3C2019decentralized}. The prefix always start with the string \texttt{did:} and is followed by a method name and a colon, e.g., the prefix for the Tangle DID method is \texttt{did:tangle:} \cite{tangle2019did}.
The NSI is a globally unique identifier, usually randomly generated, whose size and other parameters are specified by the DID method. A truncated example of a DID is: \texttt{did:tangle:WILTZRG...Q99NA9999}. Thus, the primary use for DIDs is to uniquely identify an entity in a decentralized way.

Another use for DIDs is to associate it with related metadata, such as public keys and service endpoints. This association is referred to as a DID Document \cite{W3C2019decentralized},
and it is useful since it enables remote agents to securely message a DID owner. 
DDos are usually serialized in JSON. Although a binary serialization is specified \cite{W3C2019decentralized},
none of the currently registered DID methods uses it.

\subsection{Swarm}
Swarm is a distributed collection of cooperating things \cite{costa2015swarm_tce}.
In the Swarm architecture,
IoT agents interact by exchanging messages through RESTful interfaces.
Two key aspects needed to guarantee a cooperative Swarm are agent identification and message security. 
Agents need to be uniquely identified so that they can be told apart from each other, and since the Swarm is distributed and may have trillions of devices, agent identification must be decentralized and scalable.
Messages exchanged between Swarm agents must be protected against attacks such as spoofing and information disclosure, in a network-agnostic way.

\section{Proposal}
\label{sec:proposal}

This section presents our proposal to enable self-sovereign identification and communication of IoT agents with low overhead for heterogeneous networks.

\subsection{Self-Sovereign Identification and Communication of IoT Agents}
\label{sec:architecture}

\begin{figure}[!t]
\centering
\includegraphics[width=9cm]{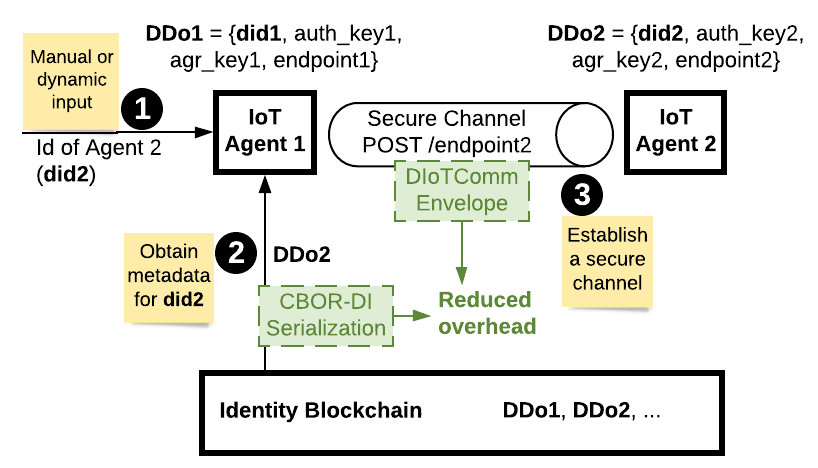}
\caption{Architecture of the proposed system.}
\label{fig:overview}
\end{figure}

Our proposal is divided into the functions of agent identification and agent communication. 
We propose agent identification as a fully self-sovereign procedure. Each agent generates its own identifier, in the format of a DID, as well as its own identity metadata, in the format of a DDo, which contains service endpoints and public keys.
This approach allows devices to fully own and control their identity without depending on third parties \cite{fedrecheski2020self}.
To enable discoverability, though, agents may choose to anchor their DDos on an Identity Blockchain, which acts as a decentralized source of truth for identity metadata.
This allows agents to dynamically resolve the DDo associated with a specific agent, given that its DID is known. For example, if Agent 1 knows the DID of Agent 2, the DDo of Agent 2 can be obtained by querying the blockchain, as shown in Figure \ref{fig:overview}.

Once agents are identified, they can begin to communicate securely. 
We consider interactions to involve an initiator agent and a receiver agent, and optionally the Identity Blockchain.
If initiator and receiver are pre-provisioned with each other's DDo,
communication can begin immediately, without the need to contact a third party.
If, on the other hand, an agent is only given the DID of another agent, it needs the blockchain to retrieve its respective DDo\footnote{Note that the DDo can be cached after the first use.}.
The latter case is shown in Figure \ref{fig:overview}.
Then, once the initiator has the DDo of the receiver, it can extract the endpoint to find out where to send the messages and use the public key to protect the messages, i.e., derive a session key for encryption.

These procedures, however, may be of limited use in constrained IoT networks due to the overhead of (1) DID resolution and (2) message protection, as discussed in Section \ref{sec:introduction}.
Existing works adopting Self-Sovereign Identity in IoT either do not consider the limitations of constrained networks, or address it by creating a centralized adaptation layer.
In this work we propose a set of extensions and optimizations to reduce the overhead of both transmitting DDos and protecting interactions between self-sovereign IoT agents. First, we propose a lean DID method that specifies the minimum needed metadata for DDos in the IoT. Then, we propose an alternative serialization mechanism for DDos, named CBOR-based DID Documents for IoT (CBOR-DI), that can reduce DDo size up to four times. Finally, we present a binary alternative to DIDComm, named DIoTComm, that significantly reduces overhead of DID-aware communications.

\subsection{The Swarm DID Method}
\label{subsec:swarm-did}

A \textit{DID method} consists of a set of definitions about the format of DIDs and DDos, as well as on how to perform management operations \cite{W3C2019decentralized}.
In this section, we present the Swarm DID Method (\texttt{did:sw:}), which will enable self-sovereign identification of IoT agents.
Although it was motivated by the Swarm architecture,
it is sufficiently generic to be used in general IoT architectures.

\subsubsection{Requirements}
Previously, we established that the self-sovereign approach can satisfy the requirements of privacy and decentralization for IoT devices \cite{fedrecheski2020self}.
We now specify the remaining requirements that need to be tackled in order to enable self-sovereign identification of devices in heterogeneous networks. 
First, both the DID and the DDo must be short since they may be carried over constrained networks.
Second, the DID should carry enough randomness to be able to identify trillions of devices.
And third, the DDo should support at least one service URL, needed to allow remote service invocations.
The next sections specify the DID and DDo according to these requirements.



\subsubsection{DID}
In the Swarm DID method, each device is responsible for autonomously generating its own DID.
A DID is composed of a prefix and a namespace-specific identifier (NSI). For the part of the prefix that identifies the DID method in use, we chose the two-letter string \texttt{sw}. Thus the full prefix is \texttt{did:sw:}.
Next, we define the NSI as a short byte array of size 16, that must be generated using a strong random number generator. The address space is $2^{128}$, what leaves more than $2^{88}$ unique identifiers for each device, considering 1 trillion devices and a uniform distribution. 
A full example of a Swarm DID with a Base58-encoded NSI is  \texttt{did:sw:TTbs19FJKYf6jXzS1dbnqe}. 

\subsubsection{DID Document}

Additional metadata about a DID can be stored in a DID Document. 
In a generic way, the DDo usually carries the DID itself, one ore more public keys, and zero or more endpoints \cite{W3C2019decentralized}.
Existing DID methods define that the DDo will contain the DID itself and at least one authentication key \cite{sovrin2020did} \cite{ockam2018did}. 
We adopt this design for the Swarm DID method, since it provides identification and authentication.
Next, unlike most DID methods, we propose the use of at least one static agreement key, since it enables the creation of a secure channel without transmission of ephemeral keys, thus saving bandwidth in constrained networks.

One consideration we take in order to shorten the DDo is that both the authentication and the agreement keys must use an optimized cipher suite with respect to public key sizes. While many elliptic curves satisfy this requirement, we adopt the curves X25519 and Ed25519 \cite{steele2019did}, which have the smallest public keys, i.e., 32 bytes.
To allow referencing specific keys within a DDo, keys may have an arbitrary identifier that is unique in the scope of the DDo. We define the following automated way to generate a short key id: compute the SHA-2 of the public key, and truncate it to the first eight bytes.

Finally, a DDo in the Swarm must support at least one service endpoint to allow remote service invocations. The main parameter for each endpoint is an URL, which enables remote agents to message the owner of the DDo. Optionally, endpoints can also have a type tag and an id that shall be unique within the DDo. Both the service type and the id are application-dependent, and if used, they should be short.

\subsection{Optimized DDo Serialization with CBOR-DI}
\label{subsec:ddo-serialization}

Serialization mechanisms have direct impact on the size of messages transferred across a network, and range from simple raw bytes encoding to complex structured data, such as the eXtensible Markup Language (XML)\footnote{https://www.w3.org/TR/2008/REC-xml-20081126/}.
While the binary approach has the benefit of conciseness, a structured approach facilitates arbitrary manipulation. Other formats, such as the JavaScript Object Notation (JSON), have provided a reasonable trade-off, with the benefit of being human-readable.
The general specification for DIDs \cite{W3C2019decentralized} uses JSON as its main format, and most existing DID methods rely on JSON as well.



\begin{figure}[!t]
\centering
\includegraphics[width=7cm]{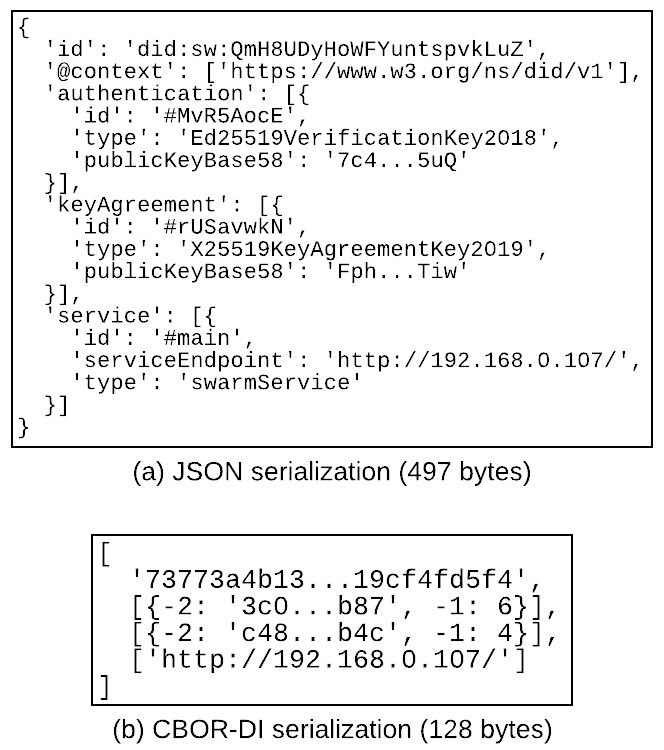}
\caption{Example of Swarm DID Document serialized in JSON and CBOR-DI.}
\label{fig:ddo-serializations}
\end{figure}

We provide a JSON-based serialization for the Swarm DID method, as shown in Figure \ref{fig:ddo-serializations} (a). It contains an identifier (DID), two public keys, and a service endpoint. The random part of the DID is serialized in Base58, since JSON does not allow encoding of raw bytes. The keys have each an id and a value, both encoded in Base58, and a type indicating its format and usage. Similarly, the service contains an id, a type, and an endpoint URL. After trimming white spaces, the JSON document occupies 497 bytes.

Although human-readable and relatively short, the JSON-based DDo still cannot be transmitted over low-power IoT networks, e.g., the LoRa\footnote{https://lora-alliance.org/resource\_hub/lorawan-specification-v1-1/} network only supports packets of up to 240 bytes. Fragmentation could be used, at the expense of increased spectrum usage and latency. What is needed is a more concise representation for DDos that allow transmission on constrained networks.

The Concise Binary Object Representation (CBOR) is a JSON-compatible serialization mechanism that uses a binary encoding. Although the DID specification considers direct conversion from JSON to CBOR \cite{W3C2019decentralized}, on average this approach only reduces document size in 20\%, i.e., achieving 415 bytes.

Considering this limitation, we present a novel serialization method named 
CBOR-based DID Documents for IoT (CBOR-DI)
that reduces size of DDos in up to 75\%.
The technique consists in transmitting only the strictly necessary parts of a DID Document, as exemplified in Figure \ref{fig:ddo-serializations} (b).
Specifically, we implement the following modifications when comparing to the JSON serialization:

\begin{itemize}
\item Use CBOR as serialization mechanism.
\item Use an array instead of a key-value mapping so that the elements have a fixed order: DID, verification keys, agreement keys, and service endpoints.
\item Remove the \texttt{did:} prefix of the DID, and only use the method designator, e.g., use \texttt{sw:} instead of \texttt{did:sw:}.
\item Encode the DID and the key values as raw bytes instead of Base58. Note that the average overhead of Base58 is close to 30\%\footnote{https://tools.ietf.org/id/draft-msporny-base58-01.html}.
\item Use the key format defined in the CBOR Object Signing and Encryption (COSE) specification \cite{RFC8152}. It defines keys as a mapping that uses integers instead of strings to reduce size. For example, instead of writing ``type: Ed25519VerificationKey2019'', we write ``-1: 6''. Also, in Figure \ref{fig:ddo-serializations} (b), the integer -2 points to the public part of the key. The full table with rules for key representation is available in the Key Objects section of COSE \cite{RFC8152}.
\item Finally, ignore optional fields in service endpoints and only include the URL.
\end{itemize}

As shown in Figure \ref{fig:ddo-serializations} (b), CBOR-DI achieves a DDo size of only 128 bytes, enabling DDo transmission in constrained networks, while losing no essential information.
Also, by leveraging existing standards, such as CBOR and COSE, it fosters interoperability.
Furthermore, the conversion process between JSON and CBOR-DI can be automated by applying a small set of mapping rules, e.g., convert between JSON and CBOR, Base58 and binary, and JSON keys and COSE keys.
Finally, although we proposed CBOR-DI in the context of the Swarm DID method, the technique is generic and could be easily extended to reduce DDo size in other methods as well.

\subsection{Secure Communication with DIoTComm}



\begin{figure}[!t]
\centering
\includegraphics[width=8cm]{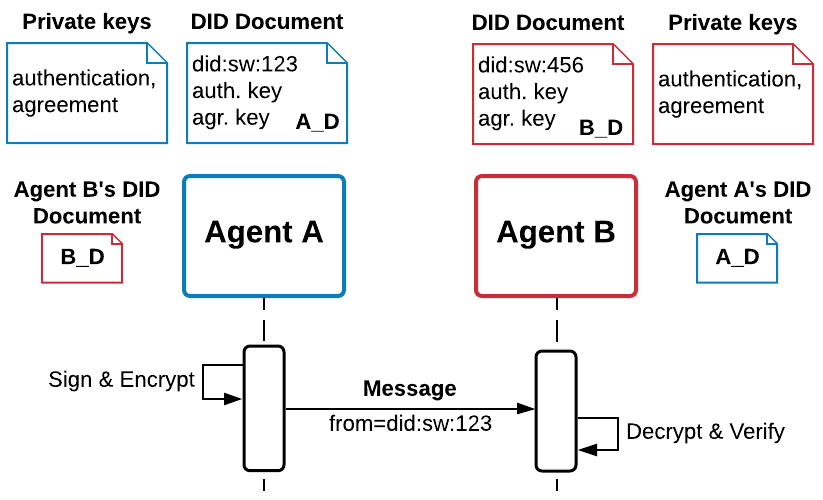}
\caption{An overview of the DIoTComm protocol.}
\label{fig:did+cose}
\end{figure}

Once two agents are identified and have access to each other's DDo, they can exchange messages securely.
The DIDComm protocol has been proposed as a way to protect communications between self-sovereign agents \cite{DIF2020didcomm}. It adopts the structure and algorithms defined in the JSON Object Signing and Encryption (JOSE) standard \cite{RFC7518}, which allows use of existing schemes for message encryption and authentication. 
Furthermore, it is agnostic of both DID method and transport protocol. 
DIDComm also defines a set of message headers to identify a message sender and receiver, as well as a message type, a unique message identifier, and other optional metadata.
One downside of DIDComm, however, is its reliance on JSON which causes overhead in constrained networks.

In this section we propose DID-based IoT Communication (DIoTComm), an alternative to DIDComm that is tailored for the IoT, i.e., it uses a more concise serialization method and simpler message headers.
While in DIDComm a JSON-based format is used for message protection, DIoTComm uses COSE, which defines both a message format and a set of lightweight security algorithms, leading to small-footprint protected messages.
In DIoTComm the only message header used is the sender id, as shown in Figure \ref{fig:did+cose}. We are able to omit the receiver id since the decryption by a receiver other than the intended one will fail. We also consider that message type and unique id, if needed, will be handled at the payload layer, e.g., if the payload is a CoAP message, its header would include a method and path, and an id.

DIoTComm leverages the structure defined by COSE messages which have one integrity-protected header, an unprotected header, a payload, and optional extra fields. We define that the sender id in DIoTComm must be binary-encoded and carried within the ``key id'' field of the protected header.
If the message must be encrypted, the payload will contain the cipher-text. If it must be signed, the payload will contain the plain-text, and the message will have the signature as a fourth parameter.
In cases where a message must be protected both for non-repudiability and confidentiality, the plain-text is first signed then encrypted.
Creation of such protected envelopes is described as follows.

For message signature, the sending agent will sign the message with its private authentication key. The receiving agent can verify the signature using the key available in the sender's DDo.
The COSE algorithm used is EDDSA, which consists in the Edwards Curve Digital Signature Algorithm that is applied over curve Ed25519 keys.

Regarding encryption, the sending agent will obtain an encryption key using its private agreement key that is locally stored, along with the public agreement key available in the receiving agent's DDo.
A similar process is then executed by the receiving agent to decrypt the message, wherein the receiving agent uses its private agreement key and the sending agent's public agreement key to obtain the decryption key.
The COSE algorithm used for key derivation is the ECDH-SS-HKDF-256, which uses an elliptic curve Diffie-Hellman with two static keys, along with a key derivation function based on SHA-256. The COSE algorithm used for content encryption is AES-CCM-16-64-128, that is the Advanced Encryption Standard in CCM mode with a 64-bit tag and a 13-bytes nonce.


\subsection{Implementation}
\label{sec:implementation}
We implemented the proposed system in the SwarmLib, a library for building Swarm agents, using the Python programming language. To construct COSE messages, we used the \texttt{cose} library. We also modified the SwarmLib and added several new routines to create DIDs and DDos, to register and resolve DDos, and to protect messages before sending them to remote agents. Routines for DDo serialization in different formats, including JSON, CBOR, and CBOR-DI were also added to the SwarmLib. We used unit tests to validate the newly added routines.
We also implemented a blockchain mock, i.e. an API to create and query DID Documents. The API supports DDos in three different formats: JSON, CBOR, and CBOR-DI. It also validates the signature during the creation of new DDos using the authentication keys from the DDos themselves, ensuring that the DDo was registered by its own agent.

\section{Evaluation}
\label{sec:evaluation}




\subsection{DID and DDo sizes}


In this section we measure the size of our proposed DID and DID Document, and compare it to five existing DID methods, as shown in Table \ref{tab:sizes}. 
The methods \texttt{did:ockam} \cite{ockam2018did}, \texttt{did:io} \cite{io2020did}, and \texttt{did:tangle} \cite{tangle2019did} were selected since they are specifically tailored for IoT applications. The other two, \texttt{did:sov} \cite{sovrin2020did} and \texttt{did:v1} \cite{v12020did}, were selected as references since they provide both a complete specification and a mature open source implementation.

As shown in Table \ref{tab:sizes}, our proposed method, \texttt{did:sw}, has the smaller DID size, occupying only 19 bytes when using the binary serialization described in Section \ref{subsec:ddo-serialization}. Among the methods tailored for IoT, \texttt{did:ockam} has the second smaller DID, requiring 39 bytes.  
In order to compare DID Documents, we built documents with equivalent configurations, i.e., having two public keys and, when applicable, one service endpoint\footnote{Some DID methods do not use endpoints.}. The measured DDos were extracted from the specification of each DID method, and a second public key was added when the example contained only one. As shown in Table \ref{tab:sizes}, the \texttt{did:sw} method has the smallest DDo size, which represents a reduction of almost 75\% when comparing to the second smallest DDo. 
These results confirm that the methods proposed in sections \ref{subsec:swarm-did} and \ref{subsec:ddo-serialization} indeed reduced DID and DDo sizes when comparing to previous works.

\begin{table}[!t]
\centering
\caption{Comparison with existing DID Methods.}
\label{tab:sizes}
\resizebox{\textwidth/2}{!}{
\begin{threeparttable}
\begin{tabular}{lllllll}
\textbf{Prefix} & \textbf{\begin{tabular}[c]{@{}l@{}}Focus\\ on IoT?\end{tabular}} & \textbf{\begin{tabular}[c]{@{}l@{}}DID\\ serialization\end{tabular}} & \textbf{DID size} & \textbf{\begin{tabular}[c]{@{}l@{}}DDo ser.\end{tabular}} & \textbf{DDo size} \\ \hline
did:sw: & Yes  & binary & \textbf{19} & Binary & \textbf{128} \\
did:sov: & No  & text & 30 & JSON & 499 \\
did:ockam: & Yes  & text & 39 & JSON & 779 \\
did:io: & Yes  & text & 49 & JSON & 1112 \\
did:v1: & No  & text & 54 & JSON & 1182 \\
did:tangle: & Yes  & text & 92 & JSON & 853 \\ \hline
\end{tabular}
\begin{tablenotes}
\end{tablenotes}
\end{threeparttable}
}
\end{table}

\begin{figure}[!t]
\centering
\begin{minipage}{0.49\textwidth}
\centering
\includegraphics[width=0.8\textwidth]{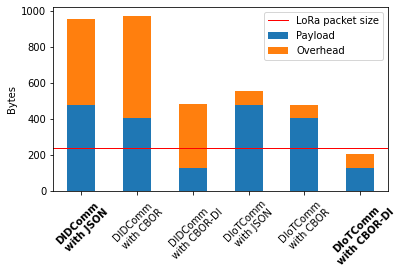}
\caption{Size and overhead for DID Documents sent within a signed message (step 2 of Figure \ref{fig:overview}).}
\label{fig:ddo_size}
\end{minipage}\hfill
\begin{minipage}{0.49\textwidth}
\centering
\includegraphics[width=0.8\textwidth]{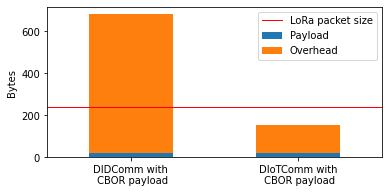}
\caption{Size and overhead for regular messages sent within a signed and encrypted message (step 3 of Figure \ref{fig:overview}).}
\label{fig:msg_size}
\end{minipage}
\end{figure}

\subsection{Secure Envelope Overhead}
In this section, we compare the overhead of using DIDComm and DIoTComm to protect DID Documents and application messages for transmission in constrained networks. 

We start by measuring the size of a signed message containing a DID Document, using both DIDComm and DIoTComm. 
In doing this, we use different DDo serializations: JSON, CBOR, and CBOR-DI. 
Figure \ref{fig:ddo_size} shows the results. The three leftmost bars use the DIDComm message envelope\footnote{The size of DIDComm with CBOR is larger than DIDComm with JSON due to overhead of Base64 encoding.}, while the three rightmost bars use DIoTComm. 
We also highlight the threshold for transmission of LoRa packets when considering Data Rate 6, which allows packets of up to 242 bytes.
Note that, although the overhead is significantly reduced when adopting DIoTComm, the only scenario in which a DDo can be transmitted over a LoRa network is when DIoTComm and CBOR-DI are combined.

In the next chart, shown in Figure \ref{fig:msg_size}, we consider a 21-bytes application message serialized in CBOR and sent over DIDComm and then over DIoTComm. Differently from the previous chart, this message is not only signed, but also encrypted, i.e. the messages are nested with two layers of headers. This confers the DIoTComm version an even higher compression rate, with an overhead at least five times smaller.

\section{Conclusion}
\label{sec:conclusion}

This paper presented a solution for self-sovereign identification and communication of IoT agents in constrained networks. 
While previous works either did not consider constrained networks, or proposed centralized solutions, we proposed a set of techniques to enable native self-sovereign identity in IoT environments.
First, we presented a specification for Decentralized Identifiers (DIDs) that focus on reduction of metadata size by using shorter identifiers and optimized cipher suites. 
We then introduced a novel serialization mechanism named CBOR-based DID Documents for IoT (CBOR-DI), which reduces DID Documents up to four times when compared to a JSON serialization.
Finally, we proposed DIoTComm, an optimized layer for protection of messages exchanged between self-sovereign agents which uses a binary encoding, thus achieving five times reduction for signed and encrypted messages.
We implemented these proposals within the Swarm framework and evaluated them with respect to metadata size and overhead. 
Regarding identity metadata, we achieved a reduction of 3.89 times when compared to related works. With respect to secure communication, we achieved a reduction of almost five times. The combination of these techniques enable the native use of self-sovereign identity in constrained IoT networks such as LoRa.
Future work includes evaluation in a real network scenario and integration with a system for authentication and authorization.

\section{Acknowledgments}

We would like to thank researchers at the Laboratory for Integrated Systems at POLI USP, the Interdisciplinary Center of Interactive Technologies at USP, and the Berkeley Wireless Research Center. Authors would also like to thank for the funding provided by LSITEC.

\ifCLASSOPTIONcaptionsoff
  \newpage
\fi

\bibliographystyle{IEEEtran}
\bibliography{main}

\end{document}